\documentclass[conference,10pt]{IEEEtran}
\usepackage{diagbox} 
\usepackage{graphicx}
\usepackage{epstopdf}
\usepackage{psfrag}
\usepackage{subfigure}
\usepackage{url}
\usepackage{stfloats}
\usepackage{amsfonts,amssymb,amsmath,bm,paralist,theorem,cite,ifthen,color,nccmath}
\usepackage{caption}
\usepackage{calc}
\usepackage{enumerate}
\usepackage{multirow}
\usepackage[ruled]{algorithm2e}

\usepackage{array}
\usepackage{float}
\usepackage{soul}
\newtheorem{proposition}{Proposition}
\newtheorem{remark}{Remark}

\graphicspath{{Figures/}}

\newcommand\Ccl{\ensuremath{\mathcal{C}}}

\newcommand\Ncl{\ensuremath{\mathcal{N}}}

\newcommand\Tcl{\ensuremath{\mathcal{T}}}

\newcommand\Ucl{\ensuremath{\mathcal{U}}}

\newcommand\Cs{\ensuremath{{\mathbb{C}}}}
\newcommand\Es{\ensuremath{{\mathbb{E}}}}

\newcommand\Bb{\ensuremath{ \mathbf{B} }}
\newcommand\Cb{\ensuremath{ \mathbf{C} }}

\newcommand\Gb{\ensuremath{ \mathbf{G} }}
\newcommand\Hb{\ensuremath{ \mathbf{H} }}
\newcommand\Ib{\ensuremath{ \mathbf{I} }}

\newcommand\ab{\ensuremath{ \mathbf{a} }}

\newcommand\eb{\ensuremath{ \mathbf{e} }}
\newcommand\gb{\ensuremath{ \mathbf{g} }}

\newcommand\hb{\ensuremath{ \mathbf{h} }}

\newcommand\ssb{\ensuremath{ \mathbf{s} }}
\newcommand\rb{\ensuremath{ \mathbf{r} }}

\newcommand\wb{\ensuremath{ \mathbf{w} }}
\newcommand\xb{\ensuremath{ \mathbf{x} }}

\newcommand\zb{\ensuremath{ \mathbf{z} }}

\newcommand\etab{\ensuremath{{\bm \eta}}}

\newcommand\diag{\ensuremath{{\rm diag}}}

\newcommand\ghz{\textrm{GHz}}
\newcommand\mhz{\textrm{MHz}}
\newcommand\dB{\textrm{dB}}

\newcommand\Nc{\ensuremath{ N_{\rm c} }}
\newcommand\Ts{\ensuremath{ T_{\rm s} }}
\newcommand\Tu{\ensuremath{ T_{\rm u} }}

\newcommand\fs{\ensuremath{ f_{\rm s} }}
\newcommand\fc{\ensuremath{ f_{\rm c} }}

\newcommand\Bw{\ensuremath{ B_{\rm w} }}

\newcommand\sigman{\ensuremath{ \sigma_{\rm n} }}

\usepackage[labelsep=period,font=footnotesize]{caption}

\IEEEoverridecommandlockouts

\begin{document}

\title{Analysis of Oversampling in Uplink Massive MIMO-OFDM with Low-Resolution ADCs}
\author{
\IEEEauthorblockN{Mengyuan Ma, Nhan Thanh Nguyen, Italo Atzeni, and Markku Juntti}
\IEEEauthorblockA{Centre for Wireless Communications (CWC), University of Oulu, Finland \\
Email: \{mengyuan.ma, nhan.nguyen, italo.atzeni, markku.juntti\}@oulu.fi
}}

\maketitle

\begin{abstract}
       Low-resolution analog-to-digital converters (ADCs) have emerged as an efficient solution for massive multiple-input multiple-output (MIMO) systems to reap high data rates with reasonable power consumption and hardware complexity. In this paper, we analyze the performance of oversampling in uplink massive MIMO orthogonal frequency-division multiplexing (MIMO-OFDM) systems with low-resolution ADCs. Considering both the temporal and spatial correlation of the quantization distortion, we derive an approximate closed-form expression of an achievable sum rate, which reveals how the oversampling ratio (OSR), the ADC resolution, and the signal-to-noise ratio (SNR) jointly affect the system performance. In particular, we demonstrate that oversampling can effectively improve the sum rate by mitigating the impact of the quantization distortion, especially at high SNR and with very low ADC resolution. Furthermore, we show that the considered low-resolution massive MIMO-OFDM system can achieve the same performance as the unquantized one when both the SNR and the OSR are sufficiently high. Numerical simulations confirm our analysis.
\end{abstract}
	
	
\section{Introduction}

Massive multiple-input multiple-output (MIMO) is a crucial physical-layer technology for current and future wireless systems \cite{Raj20}, which provides high spectral efficiency thanks to the large number of antennas at the base station (BS) \cite{albreem2019massive}. However, when massive MIMO is adopted at millimeter wave and (sub-)THz frequencies \cite{bjornson2019massive,shafie2022terahertz}, its energy efficiency can be severely burdened by the high power consumption of each radio-frequency (RF) chain. In this respect, analog-to-digital converters (ADCs) are the most power-hungry RF components, as their power consumption increases exponentially with the number of resolution bits \cite{murmann2015race}. For instance, high-speed ADCs (e.g., operating at 1~Gsample/s) with high resolution (e.g., 8--12 bits) can consume several Watts \cite{li2017channel}. Therefore, adopting low-resolution ADCs at the BS has been regarded as an effective approach to reducing power consumption without excessively compromising the performance \cite{liu2019low,Atz22}.

Despite the reduced power consumption, low-resolution ADCs introduce a non-linear quantization distortion to the signal, which cannot be eliminated by increasing the transmit power. While adding more antennas at the BS can compensate for the performance loss due to the quantization distortion \cite{fan2015uplink,mollen2016uplink,jacobsson2017throughput}, it also raises the overall power consumption and hardware complexity. On the other hand, temporal oversampling can improve the sum rate in quantized massive MIMO systems without increasing the number of antennas and RF chains \cite{uccuncu2018oversampling}. Furthermore, oversampling enables higher-order modulation over a 1-bit quantized single-antenna additive white Gaussian noise (AWGN) channel \cite{krone2012capacity}. In addition, it was shown in \cite{deng2020bandlimited} that the sum rate grows roughly logarithmically with the oversampling ratio (OSR). Most of the aforementioned studies consider narrowband or single-carrier systems and are not readily applicable to wideband multi-carrier scenarios in general. This is because the correlation of time-domain symbols due to the low-resolution ADCs makes the frequency-domain signal model for multi-carrier systems more involved \cite{jacobsson2019linear}. Massive MIMO orthogonal frequency-division multiplexing (MIMO-OFDM) systems with low-resolution ADCs and oversampling were studied by {\"U}{\c{c}}{\"u}nc{\"u} \textit{et al.} in \cite{uccuncu2019performance} under adjacent channel interference (ACI). Specifically, this work analyzed the performance with zero-forcing (ZF) combining and showed that oversampling can improve the signal-to-interference-plus-noise-and-distortion ratio (SINDR) and suppress the ACI in both 1-bit and multi-bit quantized systems.

Inspired by \cite{uccuncu2019performance}, we perform a deeper analysis of how the OSR, the ADC resolution, and the SNR collectively affect the performance of uplink massive MIMO-OFDM systems with low-resolution ADCs and oversampling, which was not reported in \cite{uccuncu2019performance}. We first present the frequency-domain signal model for an uplink MIMO-OFDM system, which accounts for the impact of low-resolution ADCs on the received time-domain symbols. We then derive an approximate closed-form expression of an achievable sum rate based on the Bussgang decomposition, which considers the temporal and spatial correlation of the quantization distortion. Our analysis reveals that oversampling can significantly improve the sum rate by mitigating the quantization distortion, especially at high SNR and with very low ADC resolution (down to 1-bit). We further demonstrate that the considered low-resolution massive MIMO-OFDM system can achieve the same performance as its unquantized counterpart when both the SNR and the OSR are sufficiently high. Numerical simulations validate our analysis and highlight the trade-off between the OSR and the ADC resolution in terms of energy efficiency and hardware complexity.

\section{System Model}

We consider an uplink massive MIMO system where a BS equipped with $M$ antennas receives signals from the $U$ single-antenna user equipments (UEs). The OFDM is assumed over a wideband channel to deal with the frequency selectivity. Specifically, let $\Delta f= \frac{1}{\Tu}$ be the subcarrier spacing, where the OFDM symbol duration $\Tu$ is assumed to be fixed. Let $f_{k}=\fc+\left(k+1-\frac{\Nc+1}{2} \right)\Delta f$, $k=0,\ldots, \Nc-1$ denote the $k$-th subcarrier frequency, where $\fc$ is the center carrier frequency. Among the total $\Nc$ subcarriers, $K$ subcarriers are employed for signal transmission, while other $\Nc-K$ subcarriers are employed for oversampling \cite{jacobsson2019linear}. Let $\tilde{s}_u[k]$ be the transmit symbol of the $u$-th UE at subcarrier $k$ with $\Es\left[|\tilde{s}_u[k]|^2\right]=1,~ k=0,\ldots, K-1$. Note that $\tilde{s}_u[k]=0$ for $k=K,\ldots,\Nc-1 $ when $\Nc >K$. Because the sampling frequency is $\fs=\Nc \Delta f$  while the transmission bandwidth of signals is $\Bw=K \Delta f$, the OSR is defined as $\beta= \frac{\Nc}{K}$. Hence, $\beta=1$ and $\beta>1$ indicate the Nyquist sampling and the oversampling scheme, respectively. The time-domain symbol is obtained by $\Nc$-points inverse discrete Fourier transform (IDFT), which can be expressed as
\begin{equation}\label{eq:IDFT of stt frequency}
    s_u[n]=\frac{\sqrt{p}}{\sqrt{\Nc}}\sum\limits_{k=0}^{\Nc-1}\tilde{s}_u[k] e^{j\frac{2\pi nk}{\Nc}}, \quad n=0,\ldots, \Nc-1,
\end{equation}
where $n$ represents the index of the time-domain symbol, and $p$ denotes the average transmit power. At the receiver, the time-domain signals are first downconverted to the baseband and transformed back to the frequency domain by $\Nc$-points discrete Fourier transform (DFT).  

Let $\ssb[n]=\left[ s_1[n], \ldots,s_{U}[n] \right]^T$ and $\tilde{\ssb}[k]=\big[ \tilde{s}_1[k], \ldots,$ \linebreak $\tilde{s}_{U}[k] \big]^T$, where $\tilde{s}_{u}[k]$ is the frequency-domain signal transmitted by the $u$-th UE, and $s_{u}[n]$ is given in \eqref{eq:IDFT of stt frequency}. The discrete-time received signal at time sample $n$ at the BS is given by
\begin{equation}\label{eq:compact form at receiver}
	\rb[n]=\sum_{d=0}^{D-1} \Hb[d] \ssb[n-d] +\wb[n], 
\end{equation}
where $D=\beta D_0$ with $D_0$ being the maximum number of delay taps under Nyquist sampling, and $\Hb[d]=\left[\hb_{1}[d],\cdots,\hb_{U}[d] \right]\in \Cs^{M\times U}$ denotes the channel matrix at the $d$-th time delay with $\hb_u[d]$ representing the channel between the $u$-th UE and the BS. Here, $ \wb[n] $ represents the AWGN vector and $\wb[n]\sim \Ccl\Ncl(\mathbf{0},\sigma^2_{\rm n}\Ib)$, where $\sigman^2$ denotes the AWGN power. By taking the DFT of both sides of \eqref{eq:compact form at receiver}, the frequency-domain received signal is expressed as
\begin{equation}\label{eq:frequency-domain unquantized model}
	\tilde{\rb}[k]= \sqrt{p} \tilde{\Hb}[k] \tilde{\ssb}[k]+\tilde{\wb}[k], \quad k=0, \ldots, \Nc-1,
\end{equation}
where $\tilde{\rb}[k]=\frac{1}{\sqrt{\Nc}} \sum_{n=0}^{N_{\rm c}-1}\rb[n] e^{-j\frac{2\pi n k}{N_{\rm c}}}$, $\tilde{\wb}[k]=\frac{1}{\sqrt{\Nc}}\sum_{n=0}^{N_{\rm c}-1}\wb[n] e^{-j\frac{2\pi nk}{N_{\rm c}}}$, and $\tilde{\Hb}[k]=\big[\tilde{\hb}_1[k],\ldots,\tilde{\hb}_U[k] \big]$ with $\tilde{\hb}_u[k]=\sum_{d=0}^{D-1}\hb_u[d] e^{-j\frac{2\pi d k}{N_{\rm c}}}$. Note that $\tilde{\rb}[k]=\tilde{\wb}[k]$ for $k=K,\ldots,\Nc-1$ as $\tilde{\ssb}[k]=\mathbf{0}$ in these cases.

\section{Signal Model with Quantization}

Assuming that the UEs employ high-resolution DACs, the BS uses identical pair of low-resolution ADCs in each RF chain for the in-phase and quadrature-phase signals. Focusing on the performance impact of ADCs, we assume in our analysis that all RF circuits other than the ADCs (e.g., local oscillators, mixers, and power amplifiers) are ideal. We further assume that the sampling rate $f_{\rm s}$ of the ADCs at the BS is the same as that of the DACs at the UE, and the system is perfectly synchronized. Finally, we assume that the spectrum of the output of ADCs is contained within $\big[-\frac{f_{\rm s}}{2},\frac{f_{\rm s}}{2}\big]$, i.e., without out-of-band emissions \cite{jacobsson2019linear}.

\subsection{Quantization Modeling}

We begin by defining the codebook of a scalar quantizer of $b$ bits as $\Ccl=\{c_0, \ldots, c_{N_{\rm q}-1}\}$, where $N_{\rm q}=2^b$ is the number of output levels of the quantizer. The quantization thresholds set is $\Tcl=\{ t_0, \ldots, t_{N_{\rm q}} \}$, where $t_0=-\infty$ and $t_{N_{\rm q}}=\infty$ allows inputs with arbitrary power. For signals with standard Gaussian distribution, the Lloyd-Max algorithm can find the optimal $\Ccl$ and $\Tcl$ that achieve the minimum square error (MSE) between the input and output of the quantizer. 
Let $Q(\cdot)$ denote the quantization function. For a complex signal $x=\Re\{x\}+j \Im\{x\}$, we have $Q(x)=Q(\Re\{x\})+j Q(\Im\{x\})$ with $Q(\Re\{x\})=c_i$ for $\Re\{x\}\in [t_i, t_{i+1}]$; $Q(\Im\{x\})$ is obtained in a similar way. When the input signal of the quantizer is a vector, $Q(\cdot)$ is applied elementwise.

The Bussgang decomposition allows to model a non-linear input-output relation of a Gaussian signal as a linear transformation \cite{bussgang1952crosscorrelation}. To model the quantization of the received signal in \eqref{eq:compact form at receiver} by the low-resolution ADCs, we first rewrite \eqref{eq:compact form at receiver} as
\begin{equation}\label{eq: signal model time domain compact}
	\Bar{\rb}= \Bar{\Hb}\Bar{\ssb} + \bar{\wb},
\end{equation}
where $\Bar{\rb}=\left[\rb[N_{\rm c}-1]^T, \ldots, \rb[0]^T \right]^T$, $\Bar{\ssb}=\big[\ssb[N_{\rm c}-1]^T, \ldots,$ \linebreak $\ssb[0]^T \big]^T$, and $\Bar{\wb}=\left[\wb[N_{\rm c}-1]^T, \ldots, \wb[0]^T \right]^T$. Furthermore, $\Bar{\Hb} \in \mathbb{C}^{M N_{\rm c} \times UN_{\rm c}}$ is a block circulant matrix \cite{uccuncu2019performance}. With the Bussgang decomposition, $\Bar{\zb}=Q(\Bar{\rb})$ can be expressed as
\begin{equation}\label{eq:ADC model}
	\Bar{\zb}=\Bar{\Bb} \Bar{\rb} + \Bar{\etab}, 
\end{equation}
where $\Bar{\zb}=\left[\zb[N_{\rm c}-1]^T, \ldots, \zb[0]^T \right]^T$, and where $\Bar{\etab}=\left[\etab[N_{\rm c}-1]^T, \ldots, \etab[0]^T \right]^T$ denotes the non-Gaussian distortion vector that is uncorrelated to $\Bar{\rb} $. Here, $\Bar{\Bb}$ represents the Bussgang gain matrix. In the case of the same resolution ($b$ bits) ADCs at all the RF chains, $\Bar{\Bb}$ reduces to a scalar $\alpha=1-\gamma$, where $\gamma$ denotes the distortion factor, satisfying $\gamma(b)=2^{-1.74b+0.28}$ for a $b$-bit ADC \cite{ma2024joint}. Therefore, \eqref{eq:ADC model} is equivalent to
\begin{equation}\label{eq:quantization model for subcarriers}
		\zb[n]=\alpha \rb[n] + \etab[n], \quad n=0,\ldots,\Nc-1. 
\end{equation}
To facilitate the performance evaluation in the frequency domain, the analysis continues by taking the DFT of both sides of \eqref{eq:quantization model for subcarriers}, yielding
\begin{align}
	\tilde{\zb}[k]&=\alpha  \tilde{\rb}[k] + \tilde{\etab}[k] \nonumber \\ 
    &=\alpha \sqrt{p} \tilde{\Hb}[k] \tilde{\ssb}[k]+\eb[k], \quad k=0, \ldots, \Nc-1,
\end{align}
where $\tilde{\zb}[k] =\frac{1}{\sqrt{\Nc}}\sum_{n=0}^{N_{\rm c}-1}\zb[n] e^{-j\frac{2\pi nk}{N_{\rm c}}} $ and $\tilde{\etab}[k] =\frac{1}{\sqrt{\Nc}}\sum_{n=0}^{N_{\rm c}-1}\etab[n] e^{-j\frac{2\pi n k}{N_{\rm c}}} $. Here, $\eb[k]\triangleq\alpha\tilde{\wb}[k]+ \tilde{\etab}[k]$ consisting of the AWGN and the quantization distortion is non-Gaussian due to $\tilde{\etab}[k]$.

\subsection{Achievable Sum Rate Analysis}

Let $\Gb[k]=\left[\gb_1[k], \ldots,\gb_U[k] \right] \in \Cs^{M\times U}$ be the combining matrix of the $k$-th subcarrier at the BS. Thus, the post-processing signal vector at subcarrier $k$ is given by
\begin{equation}
    \hat{\xb}[k]=\Gb[k]^H\tilde{\zb}[k]=\sqrt{p} \alpha \Gb[k]^H\tilde{\Hb}[k] \tilde{\ssb}[k]+\Gb[k]^H\eb[k].
\end{equation}
The $u$-th element of $\hat{\xb}[k]$ can be expressed as
	\begin{align}
	   \hat{x}_u[k]&= \underbrace{\sqrt{p}\alpha\gb_u[k]^H \tilde{\hb}_u[k]\tilde{s}_u[k]}_{\text{desired signal}} \nonumber\\
       & \phantom{=} \ + \underbrace{\sqrt{p}\alpha\sum\limits_{j\neq u}^U \gb_u[k]^H \tilde{\hb}_j[k]\tilde{s}_j[k]}_{\text{interference}} + \underbrace{\gb_u[k]^H\eb[k]}_{ 
 \substack{\text{AWGN and} \\ \text{quantization} \\ \text{distortion}}   },
    \end{align}
and the resulting SINDR is
\begin{equation}
    \zeta_u[k] = \frac{p \alpha^2 |\gb_u[k]^H\tilde{\hb}_u[k]|^2 }{p \alpha^2 \sum\limits_{j\neq u}^U |\gb_u[k]^H\tilde{\hb}_j[k]|^2+ \gb_u[k]^H\Cb_{\eb_k}\gb_u[k]},
\end{equation}
where $\Cb_{\eb_k} = \Es\left[\eb[k]\eb[k]^H\right]=\Cb_{\tilde{\etab}_k}+\alpha^2\sigman^2 \Ib$ and $\Cb_{\tilde{\etab}_k} = $ \linebreak $\Es\left[\tilde{\etab}[k]\tilde{\etab}[k]^H\right]$. Treating the interference-plus-noise-and-distortion term as a Gaussian random variable with the same variance, we obtain an achievable sum rate as \cite{fan2015uplink}
\begin{equation}\label{eq:achievable sum rate}
	R=  \sum\limits_{k=1}^K \sum\limits_{u=1}^U \Delta f \log_2\left( 1+ \zeta_u[k]\right).
\end{equation}
It is observed that $\Cb_{\tilde{\etab}_k}$ is required to compute the sum rate in \eqref{eq:achievable sum rate}. However, obtaining $\Cb_{\tilde{\etab}_k}$ may be challenging due to the quantization distortion. Alternatively, we derive an approximate closed-form expression in the following proposition.	

\vspace{-4mm}
\begin{proposition}\label{prop:Cov_eta_fd}
$\Cb_{\tilde{\etab}_k} $ can be approximated as 
\begin{equation}\label{eq:quantization noise in frequency domain}
    \Cb_{\tilde{\etab}_k} \approx \gamma(1-\gamma) \bigg(   \frac{p}{\Nc}\sum\limits_{k=0}^{K-1} \diag\left( \tilde{\Hb}[k]\tilde{\Hb}[k]^H \right) + \sigman^2\Ib\bigg),
\end{equation}
where we recall that $\gamma$ represents the inverse SQR given in \cite{max1960quantizing}, and $p$ denotes the average transmit power. The approximation in \eqref{eq:quantization noise in frequency domain} becomes more accurate at low SNR or at high SNR with a low OSR and  a high ADC resolution.
\end{proposition}
\vspace{-2mm}

\noindent Proposition \ref{prop:Cov_eta_fd} can be obtained through the DFT of the time-domain correlation matrix $\Cb_{\rb}[\iota] = \Es\left[ \rb[n]\rb[n-\iota]^H\right]$ and $\Cb_{\etab}[\iota] = \Es\left[ \etab[n]\etab[n-\iota]^H\right]$ as well as the approximation $\Cb_{\etab}[0]\approx \gamma(1-\gamma)\diag(\Cb_{\rb}[0])$ derived in \cite{mezghani2012capacity}. The detailed proof is omitted due to limited space. Note that $\Cb_{\etab}[\iota]$ includes the temporal and spatial correlations of the quantization distortion.

Using the result in \eqref{eq:quantization noise in frequency domain}, we can approximate $\Cb_{\eb_k}$ as
\begin{equation}\label{eq:Var equivalent noise}
    \Cb_{\eb_k}\approx \left(1-\gamma  \right)\left( \frac{\gamma p}{\Nc}\sum\limits_{k=0}^{K-1} \diag\left( \tilde{\Hb}[k]\tilde{\Hb}[k]^H \right)  +  \sigman^2 \Ib   \right). 
\end{equation}
From \eqref{eq:Var equivalent noise}, the SINDR can be rewritten as
\begin{equation}\label{eq:effective snr}
    \zeta_u[k]\approx \frac{ |\gb_u[k]^H\tilde{\hb}_u[k]|^2 }{\sum\limits_{j\neq u}^U |\gb_u[k]^H\tilde{\hb}_j[k]|^2+ \gb_u[k]^H\Cb_{\eb}\gb_u[k]},
\end{equation}
where
\begin{equation}\label{eq:effective noise}
    \Cb_{\eb} = \frac{\gamma}{(1-\gamma)\beta}\Hb_{\rm e} + \frac{1}{\rho(1-\gamma)}\Ib
\end{equation}
with $\Hb_{\rm e} = \frac{1}{K}\sum_{k=0}^{K-1} \diag\left( \tilde{\Hb}[k]\tilde{\Hb}[k]^H \right)$ and $\rho = \frac{p}{\sigman^2}$. Note that $\rho$ denotes the SNR. We observe that the sum rate in \eqref{eq:achievable sum rate} based on \eqref{eq:effective snr} is jointly affected by three factors, i.e., the OSR, the ADC resolution, and the SNR. We note some important observations in the following:
\begin{enumerate}
    \item With high ADC resolution, we have $\gamma \rightarrow 0$, which yields
    \begin{equation}\label{eq:snr in unquantized systems}
        \zeta_u[k] \rightarrow \frac{ |\gb_u[k]^H\tilde{\hb}_u[k]|^2 }{\sum\limits_{j\neq u}^U |\gb_u[k]^H\tilde{\hb}_j[k]|^2 + \frac{1}{\rho} \|\gb_u[k] \|^2   }.
    \end{equation}    
    Based on \eqref{eq:snr in unquantized systems}, we can readily obtain the sum rate corresponding to the unquantized system.
    \item  It can be observed that increasing the OSR helps to mitigate the quantization distortion, which results in a higher sum rate. In particular, when the OSR increases without bound, i.e., $\beta \rightarrow \infty$, $\zeta_u[k]$ is limited by the SNR, which is
    \begin{equation}\label{eq:OSR infinity bound}
        \zeta_u[k] \rightarrow \frac{ |\gb_u[k]^H\tilde{\hb}_u[k]|^2 }{\sum\limits_{j\neq u}^U |\gb_u[k]^H\tilde{\hb}_j[k]|^2 + \frac{1}{\rho(1-\gamma)} \|\gb_u[k] \|^2}.
    \end{equation}
    This implies that, as the sum rate approaches the upper bound constrained by the SNR, the advantages gained from increasing the OSR become less significant.

    \item In addition, at high SNR, oversampling can effectively improve the sum rate, especially with very low ADC resolution. In particular, when the SNR approaches infinity, i.e., $\rho \rightarrow \infty$, the second term of \eqref{eq:effective noise} approaches zero and $ \Cb_{\eb}\rightarrow  \frac{\gamma}{(1-\gamma)\beta}\Hb_{\rm e}$. Hence, we obtain
    \begin{equation}\label{eq:snr infty}
        \hspace{-7mm}
        \zeta_u[k] \rightarrow \frac{ |\gb_u[k]^H\tilde{\hb}_u[k]|^2 }{\sum\limits_{j\neq u}^U |\gb_u[k]^H\tilde{\hb}_j[k]|^2 + \frac{\gamma}{\beta(1-\gamma)} \gb_u[k]^H \Hb_{\rm e} \gb_u[k]   },
    \end{equation}
    which is limited by the quantization distortion and can be enhanced by increasing the OSR. Moreover, it is seen that a lower ADC resolution yields a larger $\frac{\gamma}{\beta(1-\gamma)}$, resulting in more significant performance enhancement due to oversampling. This is because $\gamma$ is inversely proportional to the resolution and $\frac{\gamma}{1-\gamma}$ monotonically increases with $\gamma$. On the other hand, at low SNR, the benefits of increasing the OSR can be marginal because the second term of \eqref{eq:effective noise}, i.e., the AWGN, could outweigh the quantization distortion. 

    \item When $\rho \rightarrow \infty$ and $\beta \rightarrow \infty$, we have
    \begin{equation}\label{eq: effective snr upper bound}       
        \zeta_u[k] \rightarrow \frac{ |\gb_u[k]^H\tilde{\hb}_u[k]|^2 }{\sum\limits_{j\neq u}^U |\gb_u[k]^H\tilde{\hb}_j[k]|^2  },
    \end{equation}
    which yields an upper bound of \eqref{eq:snr in unquantized systems} when $\rho \rightarrow \infty$. This implies that, with sufficiently large SNR and OSR, a 1-bit quantized system can perform similarly to its unquantized counterpart.
\end{enumerate}
We summarize the above discussions in the following remark:

\vspace{-4mm}
\begin{remark}
Oversampling can effectively improve the sum rate of low-resolution systems, especially at high SNR. In general, oversampling performs better at higher SNR and with lower ADC resolution. In particular, when both the SNR and the OSR are sufficiently large, the performance of the quantized system approaches that of the unquantized one.
\end{remark}
\vspace{-2mm}

\section{Simulation Results}\label{sec:simulations}

We consider the maximum ratio combining (MRC) to evaluate the achievable sum rate, which is $\Gb[k]= \Hb[k]\diag\left(\Hb[k]^H\Hb[k] \right)^{-1}$. The delay-$d$ channel between the $u$-th UE and the BS is modeled as \cite{alkhateeb2016frequency}
\begin{equation}
    \hb_u[d]=\sqrt{\frac{M}{L}}\sum\limits_{\ell=1}^{L} \beta_{u,\ell} p(d\Ts-\tau_{u,\ell})\ab(\theta_{u,\ell}),
\end{equation}
where $\beta_{u,\ell}$, $\tau_{u,\ell}$, and $\theta_{u,\ell} $ denote the $\ell$-th path gain, path delay and angle-of-arrival, respectively. Here, $ p(t) $ represents the pulse-shaping function following the same parameters as in \cite{alkhateeb2016frequency}. In the simulations, we assume $\beta_{u,\ell}\sim\Ccl\Ncl(0,1)$, $\tau_{u,\ell}\sim \Ucl \big[0,\frac{D_0}{\Bw} \big]$ with $D_0=\frac{K}{4}$ as in \cite{park2017dynamic}, and $\theta_{u,\ell}\sim \Ucl[0,2\pi]$. Here, $\Ucl[a,b]$ represents the uniform distribution in the interval $[a, b]$. The array steering vector is expressed as $\ab(\theta)=\frac{1}{\sqrt{M}}\left[1,e^{-j\pi \sin(\theta)},\ldots, e^{-j(M-1)\pi \sin(\theta)}\right]$. Furthermore, we set $\fc=140~\ghz$, $\Delta f=10~\mhz$, $K=128$, and $L=3$ due to the channel sparsity in the (sub-)THz band. The AWGN power is $\sigman^2 = N_0 \Delta f$ with $N_0$ being the AWGN power density. The following results are obtained by averaging over $10^3$ independent channel realizations.

    \begin{figure*}[t]
		\centering
		\subfigure[1-bit ADCs.]
		{\label{fig:SE vs SNR 1bit}\includegraphics[scale=0.4]{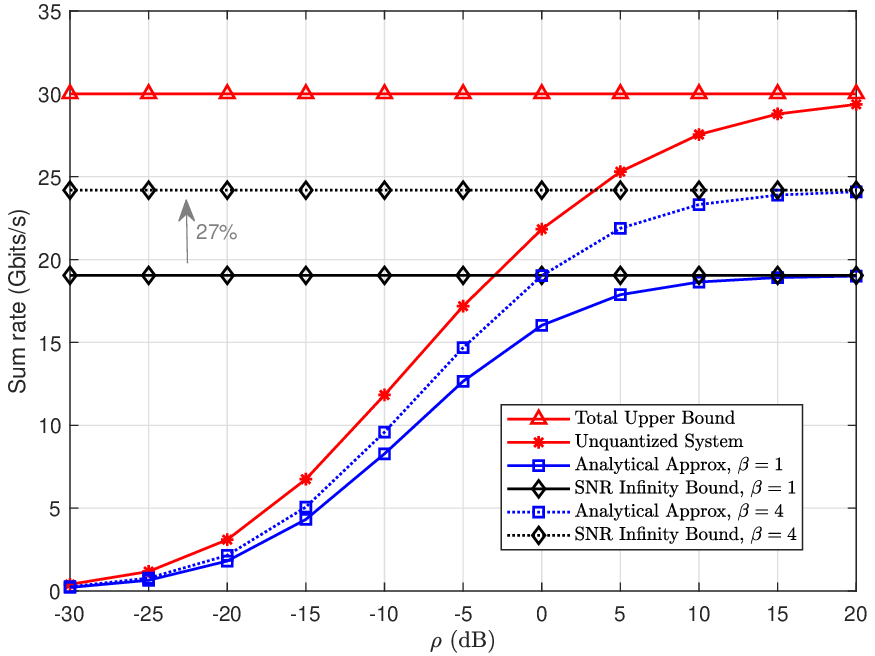}}
		\hspace{-1mm}
		\subfigure[ 2-bit ADCs.]
		{\label{fig:SE vs SNR 2bit}\includegraphics[scale=0.4]{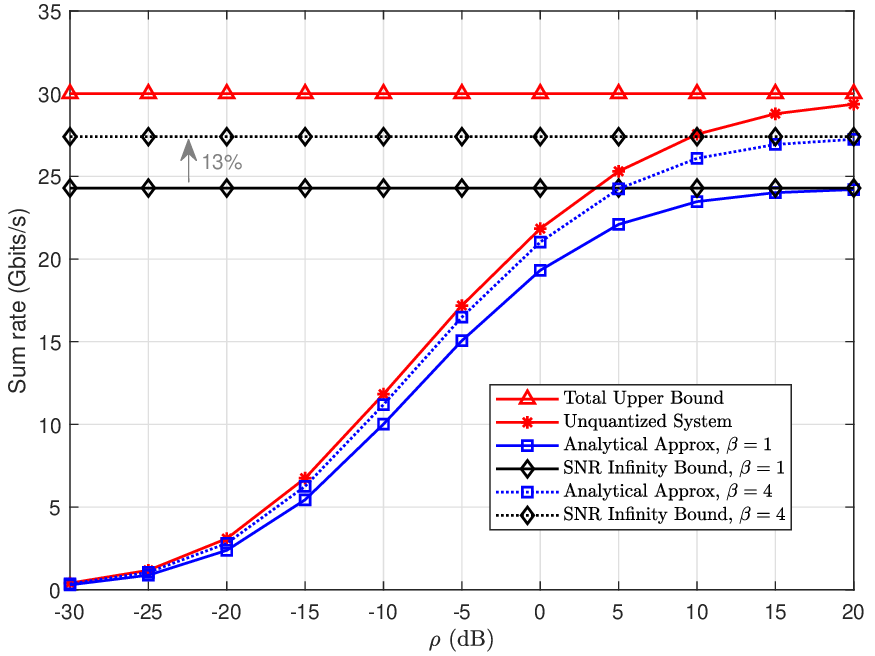}}
		\hspace{-1mm}
		\subfigure[3-bit ADCs.]
		{\label{fig:SE vs SNR 3bit}\includegraphics[scale=0.4]{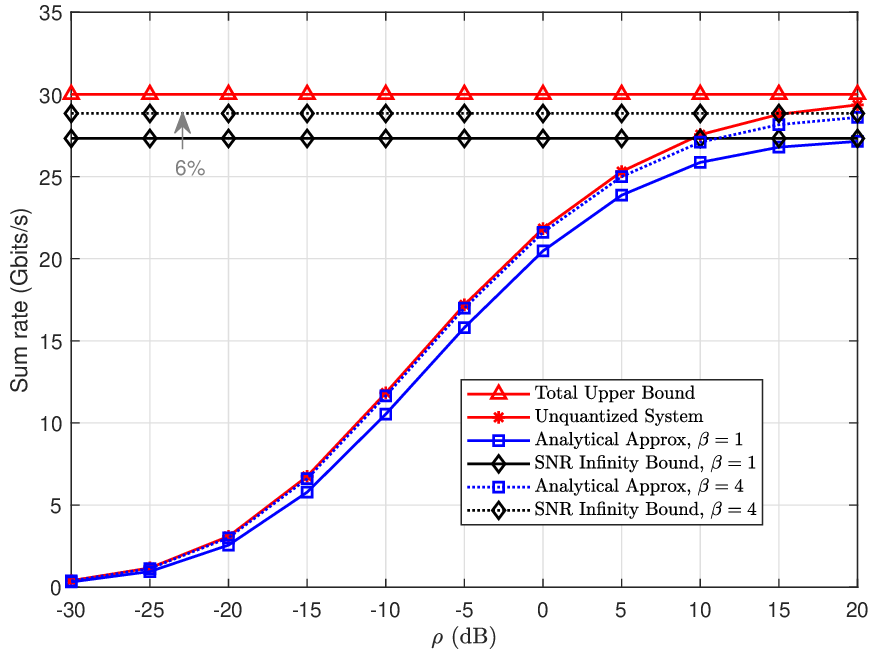}}
		\vspace{-2mm}
		\caption{Achievable sum rate versus SNR with $M=64$, $U=4$, and $K=128$.}
		\label{fig:SE vs SNR}
		\vspace{-2mm}
	\end{figure*}

    \begin{figure}[t]
    \vspace{-3mm}
        \centering
        \includegraphics[scale=0.45]{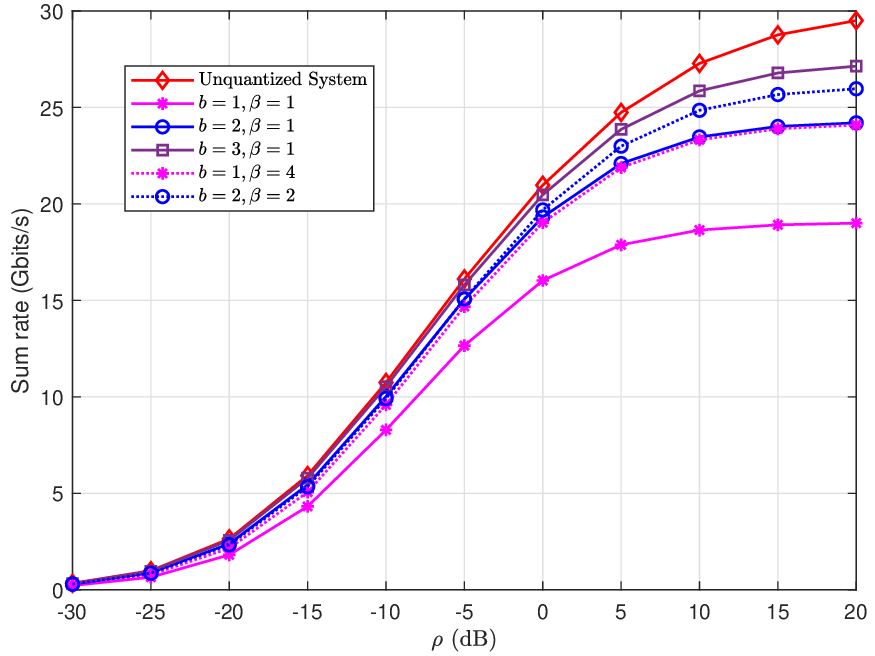}
        \vspace{-1mm}
        \caption{Achievable sum rate versus SNR with $M=64$, $U=4$, and $K=128$.}
        \label{fig:SE vs SNR cmp}
    \end{figure}

    \begin{figure}[t]
    \vspace{-3mm}
        \centering
        \includegraphics[scale=0.45]{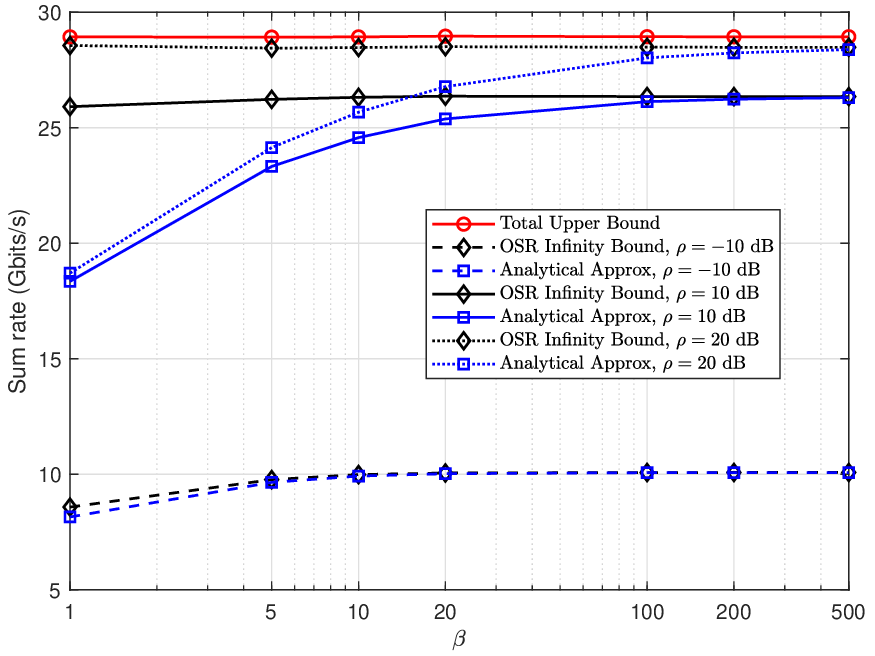}
        \vspace{-1mm}
        \caption{Achievable sum rate versus OSR with 1-bit ADCs, $M=64$, $U=4$, and $K=128$.}
        \label{fig:SE vs OSR}
    \end{figure}

Fig.~\ref{fig:SE vs SNR} shows the achievable sum rate versus the SNR with 1-bit, 2-bit, and 3-bit ADCs. Specifically, we consider: (a) the sum rate obtained with $\rho \rightarrow \infty$ and $\beta \rightarrow \infty$ in \eqref{eq: effective snr upper bound} (``Total Upper Bound''); (b) the sum rate of the unquantized system in \eqref{eq:snr in unquantized systems} (``Unquantized System''); (c) the approximate sum rate in \eqref{eq:effective snr} (``Analytical Approx''); and (d) the sum rate in \eqref{eq:snr infty} obtained with $\rho \rightarrow \infty$ (``SNR Infinity Bound''). We make the following observations from these figures. First, increasing the ADC resolution leads to significant performance improvement, and employing 3-bit ADCs allows to approach the sum rate of the unquantized system. This agrees with the findings in \cite{jacobsson2017throughput}. Second, increasing the OSR substantially enhances the sum rate, especially at high SNR and with 1-bit ADCs, e.g., $27\%$ and $6\%$ performance improvement at $\rho=20~\dB$ for the 1-bit and 3-bit quantized system, respectively. However, the performance improvement is marginal at low SNR cases due to the large AWGN. Third, it can be observed from Figs. \ref{fig:SE vs SNR 1bit} and \ref{fig:SE vs SNR 2bit} that the 1-bit quantized system with $\beta=4$ can achieve a comparable sum rate to the configuration with 2-bit ADCs. However, since the typical power consumption of ADCs can be modeled as $\kappa\fs2^b$ with $\kappa$ being the constant associated with the ADC quality \cite{murmann2015race}, there is a trade-off between the OSR and the ADC resolution in terms of energy efficiency and hardware complexity.

Fig.~\ref{fig:SE vs SNR cmp} plots the achievable sum rate versus the SNR with different OSRs and ADC resolutions, where the power consumptions of ADCs are equated for three configurations based on $\kappa\fs2^b$, namely: (i) $b=1$ and $\beta=4$; (ii) $b=2$ and $\beta=2$; and (iii) $b=3$ and $\beta=1$. The results reveal that increasing the resolution of ADCs is more effective than increasing the OSR. Specifically, the configuration with $b=3$ and $\beta=1$ achieves the highest performance. Moreover, it is observed that the system that employs 1-bit ADCs oversampled by a factor of $4$ can attain comparable performance to the 2-bit system without oversampling. However, this comes at the cost of double the energy expenditure. This observation is consistent with the results reported in \cite{uccuncu2019performance} regarding the bit error rate. Nonetheless, we remark that 1-bit quantized systems have very low hardware complexity (e.g., the automatic gain control used in multi-bit quantized systems is no longer needed), which is not taken into account in our numerical results.

Fig.~\ref{fig:SE vs OSR} depicts the achievable sum rate versus the OSR with 1-bit ADCs. The sum rate obtained with \eqref{eq:OSR infinity bound} at $\beta \rightarrow \infty$ is referred to as ``OSR Infinity Bound''. It is seen that oversampling can substantially improve the sum rate at medium-to-high SNR, while it only yields minor benefits for low SNR scenarios. Furthermore, increasing the SNR and the OSR can progressively bridge the performance gap between the 1-bit quantized and the unquantized systems. This is due to the system performance being jointly corrupted by the AWGN and the quantization distortion. While improving the SNR can overcome the AWGN, the resulting reduced randomness makes the quantization distortion more significant. Therefore, oversampling, which mitigates the quantization distortion, can further enhance the system performance. As such, oversampling is more effective at high SNR, as observed in Fig.~\ref{fig:SE vs OSR}. Ideally, when $\beta \rightarrow \infty$, the quantization distortion can be entirely suppressed, and the performance is upper-bounded by the SNR. However, it is observed that, for $\beta \geq 20$, increasing the OSR yields only minor gains. Therefore, determining a reasonable OSR is crucial to achieve a suitable trade-off between sum rate and energy efficiency, considering that the power consumption of the ADCs increases linearly with the sampling frequency.

\section{Conclusion}

We analyzed the impact of oversampling on the achievable sum rate in an uplink massive MIMO-OFDM system with low-resolution ADCs. Both the analytical and numerical results demonstrated that oversampling can significantly improve the sum rate of quantized systems by mitigating the quantization distortion. In particular, oversampling gives higher gains at higher SNR and with lower ADC resolution (especially down to 1-bit). Furthermore, we showed that the system with low-resolution ADCs can approach the performance of the unquantized system when both the SNR and the OSR are sufficiently high. Moreover, the results indicate the necessity to strike a balance between the OSR and the ADC resolution in terms of energy efficiency and hardware complexity. We note that, although these results are obtained assuming single-antenna UEs, they can be similarly derived for the scenario with multi-antenna UEs. Furthermore, we observed similar results using ZF combining, with the difference being that higher sum rate improvements are achieved due to oversampling compared to employing MRC. Future research may investigate the trade-off between the OSR and the ADC resolution to maximize the energy efficiency.

\section{Acknowledgements}

This work was supported by the Academy of Finland (332362 EERA, 336449 Profi6, 346208 6G~Flagship, 348396 HIGH-6G, and 357504 EETCAMD), Infotech Oulu, and the European Commission (101095759 Hexa-X-II).

\bibliographystyle{IEEEtran}
\bibliography{conf_short,jour_short,refs-my}

\end{document}